\documentclass[12pt]{iopart}

\usepackage{iopams}

\expandafter\let\csname equation*\endcsname\relax

\expandafter\let\csname endequation*\endcsname\relax

\usepackage{amsmath}

\usepackage{graphicx}
\usepackage{cite}
\usepackage[colorlinks, citecolor = blue, urlcolor = blue]{hyperref}

\begin{document}

\title{Superconducting properties of doped blue phosphorene: Effects of non-adiabatic approach }

\author{Mohammad Alidoosti$^1$, Davoud Nasr Esfahani$^{1,2}$ and Reza Asgari$^{3,4}$ }
\address{$^1$ Pasargad Institute for Advanced  Innovative Solutions (PIAIS), Tehran 19916-33361, Iran}
\address{$^2$ Department of converging technologies, Khatam University, Tehran 19916-33357, Iran}
\address{$^3$ School of Physics, Institute for Research in Fundamental Sciences (IPM), Tehran 19395-5531, Iran}
\address{$^4$ School   of  Physics,  University  of  New  South  Wales,  Kensington,  NSW  2052,  Australia}
\ead{r.asgari@unsw.edu.au, d.nasr@khatam.ac.ir}

\begin{indented}
\item[]\today
\end{indented}
\begin{abstract}

We study the effects of Kohn anomalies on the superconducting properties in electron- and hole-doped cases of  monolayer blue phosphorene, considering both adiabatic and non-adiabatic phonon dispersions  using first-principles calculations. We show that the topology of the Fermi surface is crucial for the formation of Kohn anomalies of doped blue phosphorene. By using the anisotropic Eliashberg formalism, we further carefully consider the temperature dependence of the non-adiabatic phonon dispersions. In cases of low hole densities, strong electron-phonon coupling leads to a maximum critical temperature of $T_c=97$ K for superconductivity. In electron-doped regimes, on the other hand, a maximum superconducting critical temperature of $T_c=38$ K is reached at a doping level that includes a Lifshitz transition point. Furthermore, our results indicate that the most prominent component of electron-phonon coupling arises from the coupling between an in-plane (out-of-plane) deformation and in-plane (out-of-plane) electronic states of the electron (hole) type doping.
\end{abstract}
\vspace{2pc}

%
%
\submitto{\TDM}
%
\maketitle
%
%

\section{Introduction}

Recent advances in the fabrication of two-dimensional (2D) materials reveal many unprecedented phenomena for further investigation of the various properties of these structures due to excellent electronic and phononic band energies~\cite{PhysRevB.91.235419,doi:10.1063/1.3665183,doi:10.1021/nl502059f,Katzmarek_2022,doi:10.1021/acs.nanolett.9b00518,PhysRevLett.114.236602,doi:10.1021/jz3012436,radisavljevic2011single}. One of the distinctive features is the superconducting properties of atomically thin 2D systems~\cite{saito2016highly,Guzman_2014,uchihashi2016two,hasan2015topological,neto2001charge,doi:10.1021/acs.jpcc.8b03108,doi:10.1021/acs.nanolett.0c05125,doi:10.1021/jp4041138,bekaert2019hydrogen,ge2015strain}. According to the BCS theory~\cite{PhysRev.108.1175} the higher superconducting transition temperature ($T_c$) can be achieved by having a remarkable electronic density of states at the Fermi level, $N(\varepsilon_{F}$),  with a large proportion of the Debye frequency, $\omega_{D}$. On the other hand, to get a significant $T_c$, one has to reach a tangible electron-phonon coupling (EPC) parameter $\lambda=2N(\varepsilon_{F}) \langle{{\bf g}}^2\rangle / \omega_D$. Nevertheless, a significant $\lambda$ arising from a large value of $N(\varepsilon_{ F})$ and $ \langle{{\bf g}}^2\rangle$, an average of EPC matrix elements over the  first Brillouin zone (1BZ), is indeed more favorable than a value of $\lambda$ arising from a small $\omega_D$. The main reason for this is that a small $\omega_D$ gives rise to a smaller 
$\omega_{ {log}}$, characteristic phonon energy, and subsequently suppresses the $T_c$.
  
Based on these analyses, one can look for light materials with a large $\lambda$ to achieve reasonable $T_c$. In the context of 2D superconductivity, there are several materials with a sharp peak in their $N(\varepsilon$) spectrum near the valence band maximum.
Examples of such systems are 2D buckled structures based on nitrogen group elements such as blue phosphorene (BLP), which is the lightest in this material group~\cite{wu2015nine,doi:10.1021/acs.nanolett.6b01459},   
and 2D $M_2X_2$ with ($M$ = In, Ga and $X$ = S and Se)~\cite{alidoosti2020charge,PhysRevLett.123.176401,PhysRevB.101.035201,zolyomi2013band,PhysRevB.89.205416,pandey2017ab}. 
For systems with large $N(\varepsilon_{F})$, there is often the possibility that a large nesting, $\xi$, occurs at some specific phonon wave vectors ($\bf q$s) leading to a considerable reduction in the bare charge susceptibility, $\chi_0$, when the temperature reduces. This character leads to an intense softening at specific $\bf q$s in some branches of the phonon spectrum, known as the Kohn anomaly~\cite{kohn1959image,doi:10.1080/23746149.2017.1343098}, and in some cases, a charge density wave (CDW) instability will appear in the form of the imaginary modes for some particular phonon wave vectors ($\textbf{q}_{{CDW}}$) when the temperature is below $T_{  {CDW}}$. Accordingly, access to such a high $T_c$ can be prohibited by this instability~\cite{alidoosti2020charge,doi:10.1021/acs.nanolett.8b00237}. Therefore, 
a more uniform distribution of $\lambda$ in terms of phonon wave vectors over the 1BZ is more favorable to gain large $T_c$, while, the formation of the CDW phase becomes less likely.

For systems where the Kohn anomaly exists, one can further include the effects of phonon spectrum normalization on temperature variations which are applied to the Eliashberg function ($\alpha^2\bf F $). This can lead to a temperature-dependent $\lambda$, so the presence of the Kohn anomaly can alter $\lambda$ and $\omega_{ {log}}$ simultaneously. 
Precisely, it is also important to consider the non-adiabatic renormalization of the phonon spectrum, leading to a more accurate determination of the formation of either the superconducting or the CDW phase at low temperatures. 
 The non-adiabatic phonon dispersions can be naturally obtained by diagonalizing the phonon dynamical matrix related to non-adiabatic non-self-consistent force constants at a physical temperature. Therefore, a self-consistent solution should be employed to achieve a better $T_c$~\cite{Note3}.

In this paper, based on first-principles calculations, we investigate the superconducting properties of BLP for some electron (N)- and hole (P)-type dopings by utilizing both adiabatic (A) and non-adiabatic (NA) approaches having been exploited in three approaches; modified isotropic Allen-Dynes~\cite{allen1975transition}, isotropic Migdal-Eliashberg (ME)  and anisotropic ME theory~\cite{migdal1958interaction,eliashberg1960interactions,PhysRevB.87.024505}.  First of all, the topology of the Fermi surface (FS) is investigated in terms of various charge dopings. Then we focus on phonon dispersion and CDW instability along with the origin of such instability at a given temperature. Furthermore, we determine that the out-of-plane phonons are strongly coupled to the electrons in P-doped cases, while, in-plane deformations have the largest contribution to the electron-phonon coupling for N-doped cases. Ultimately, we calculate the $T_c$  and  CDW phase transition temperature, $T_{ {CDW}}$, in both adiabatic  and non-adiabatic  regimes for different charge carrier densities. In this case, to better align the isotropic Migdal-Eliashberg and Allen-Dynes mechanisms, we adopted the computational implementation of potential $\mu^*_N$ rather than $\mu^*_c$ as explained in supplementary materials. Moreover, the direction-dependent $\alpha^2 \bf F$ was extracted to reveal which types of deformations have the greatest contribution to the charge carrier-phonon coupling (for both the electron and hole cases) and subsequently for T$_c$.  Our results show that the presence of non-adiabatic effects is an obstacle to entering into the unstable CDW phase for some P-doped levels. The maximum $T_c$s are calculated at around 97 and 91 K for P-doping $+0.03$ and $+0.02$ in units of electron per formula unit ($e/$f.u.), respectively.

  It should be noted that literature has primarily discussed intercalated 2D systems~\cite{durajski2021stability,huang2015prediction,PhysRevLett.27.402,ichinokura2016superconducting,zhu2018superconductivity} which require numerous atoms, however, our result demonstrates that we are able to obtain a large critical temperature at a doping level that is easily accessible in  the experiment. 

we would like to emphasize not only does our study provide a platform for understanding the fundamental physics underlying the superconductivity beyond the standard adiabatic model, but it also, and perhaps more importantly, offers an experimental route for investigating the hole doping in blue phosphorene at accessible densities. The nature of symmetry in 2D materials is what sets blue phosphorene apart from other well-known 2D materials (like graphene), as it significantly enforces in-plane electron-phonon coupling in those materials~\cite{PhysRevB.106.045301}. In contrast, blue phosphorene's buckling structure (lack of $\sigma_h$ symmetry) makes spectacular results stemming from out-of-plane deformations. Additionally, such buckling results in electron-phonon interaction strengths that are far higher than those of other 2D materials, for example, compared to monolayer InSe, graphene and TMDs. Furthermore, it should be noted that non-adiabatic effects are applied to the entire Brillouin Zone to explore the competition between the superconducting and CDW phases. 

The paper is organized as follows. In Sec.~\ref{Theory} we present, briefly, utilized  methods and computational details. In Sec.~\ref{sec:results} we describe our results about CDW formation, Kohn anomaly, and superconducting properties. Ultimately, we present our conclusions in Sec.~\ref{sec:conclusion}.


\section{Methods}\label{Theory}
First-principles calculations are performed within the framework of density functional theory (DFT), which is implemented in \textsc{quantum espresso} distribution~\cite{0953-8984-21-39-395502}. All our calculations are performed using the norm-preserving pseudopotential (Perdew-Burke-Ernzerhof)~\cite{PhysRevLett.77.3865}. The kinetic cutoff 80 Ry is used for the Kohn-Sham wave functions. A $24\times24\times1$ $\bf{k}$ point grid is used for electronic integration with the Monkhorst pack grid \cite{PhysRevB.13.5188}. In addition, a vacuum space of 20 \AA \; is used to eliminate interfering interactions between adjacent layers. Density functional perturbation theory (DFPT) \cite{baroni2001phonons} is applied to calculate the phonon frequencies and electron-phonon matrix elements of the system. In order to increase the accuracy of electron-phonon calculations, Wannier functions \cite{marzari1997n,souza2001maximally,mostofi2008wannier90} are used, implemented in the \textsc{epw} code~\cite{ponce2016epw}.
In this code, a coarse grid $24\times 24\times 1$ $\bf{k}$ point and $12\times 12\times1$ $\bf{q}$ point are first assumed to calculate the electron band structure and phonon dispersion and then a fine grid, $240\times 240\times 1$ $\bf{k}$ point and $120\times 120\times1$ $\textbf{q}$ point are used to calculate $T_c$. To evaluate the corresponding properties of the electron-phonon interaction (EPI), the Dirac delta functions are replaced by a Gaussian function with $\sigma_{el}=10$ meV and $\sigma_{ {ph}} =$ 0.2 meV for electronic and phononic integrations, respectively. In contrast, the Fermi-Dirac smearing~\cite{PhysRevB.65.035111} around 0.01 Ry is assumed as a physical parameter to explain the temperature dependence of the electronic band structure. Further details on theory and calculation methods can be found in the supplementary section.



\section{Results}\label{sec:results}

The optimized buckled hexagonal structure of pristine single-layer BLP with $D_{3d}$ symmetry has a lattice constant $a= 3.28$ {\AA} including a buckling height of about 1.24 {\AA} 
 agrees well with other reports~\cite{PhysRevLett.112.176802,doi:10.1021/acs.nanolett.6b01459,doi:10.1002/qua.26230}. 
In this work, the jellium model is adopted to account for both N- and P-doped cases. We will discuss some special N-dopings  $-0.05,\, -0.1$ and $-0.15$ $e$/f.u. related to the electron densities, $5.32\times10^{13},\, 1.05\times10^{14} $ and $1.5\times10^{14} \,$ cm$^{-2}$ respectively,  and $+0.01, \, +0.02$  and $+0.03$  $e$/f.u.  for P-doped cases tantamount to $1.07\times10^{13}$, $ 2.15\times10^{13}$ and $3.22\times10^{13} $  cm$^{-2}$ charge carrier densities. Henceforth, $\pm$ will be used to represent P or N type dopings and $e$/f.u. units related to the above doping levels are  eliminated.

\begin{table}[!]
\caption{Superconducting properties of the single-layer BLP containing the density of state at the Fermi level, $N(\varepsilon_{F}$), EPC constant, $\lambda_{ {tot}}$,  average logarithmic phonon frequency, $\omega_{ {log}}$ in unit of K, superconducting critical temperature,  $T_c$, extracted from modified  Allen-Dynes approach,  transition temperature to the CDW phase, $T_{ {CDW}}$, in both  adiabatic (A) and non-adiabatic (NA) regimes for  different charge  carrier densities.  All temperatures are in unit of K.
\label{table1}}
\begin{indented}  
\item[]   
\begin{tabular}{@{}lccccccccc}     
\hline \hline 
 Doping & $N(\varepsilon_{F}$)& $  {{\lambda}_{ {tot}}^{ {A}}} $  & $  {{\lambda}_{ {tot}}^{ {NA}}} $   & $\omega_{ {log}}^{ {A}}$  & $\omega_{ {log}}^{ {NA}}$ &$T_c^{ {A}} $ &$T_c^{ {NA}} $&T$^A_{ {CDW}} $&T$^{ {NA}}_{ {CDW}} $    \\
 
\hline

\bf+0.03 \;  & 2.12  \;& 5.16 \; & 5.30 \; &232 \; & 228 \; & 81\; & 81 & 300& $<$2\\
\bf+0.02 \;  & 1.84  \;& 4.42 \; & 4.47 \; & 240 \; & 239 \; & 76\; & 76& 210& $<$2 \\
\bf+0.01 \;  & 1.35 \;&  3.78 \; &3.77 \; & 219 \; & 230  \; & 65\; & 67& 105& $<$2 \\
\bf-0.05 \;  & 0.37  \;& 0.76 \; & 0.84 \; & 220 \; & 177 \; & 9\; & 9& $<$2& $<$2  \\
\bf-0.10 \;  & 0.68  \;&  2.05 \; &2.11 \; & 137 \; & 139  \; &22\; & 26& $<$2& $<$2   \\
\bf-0.15 \;  & 0.55  \;&  1.55 \; &1.66 \; &139 \; & 141 \; &19\; & 21& $<$2& $<$2 \\
\hline \hline
\end{tabular} 
\end{indented}    
\end{table}

\begin{figure}[!]
\centering
\includegraphics[scale=0.685, trim={-2cm 1cm 0cm 0cm }]{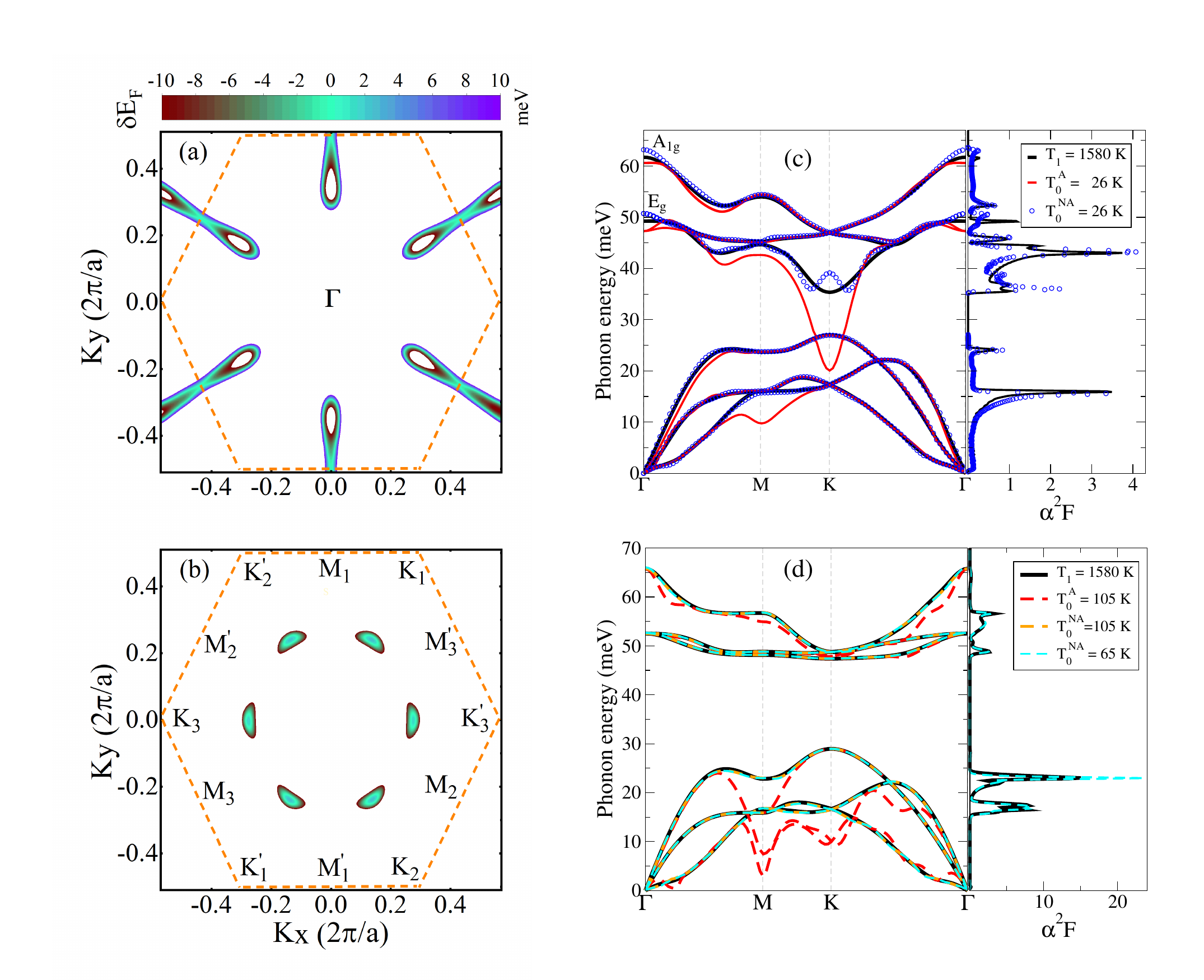}
\caption{ The  FS  contour and phonon dispersion of the  single-layer BLP based on
the jellium model. (a) and (b) depicted  the  FS  for -0.1 and  +0.01,  respectively. Orange dashed lines are applied to display 1BZ boundaries. 
(c) Phonon dispersion with respect to electronic temperatures for the case $-$0.1 by considering both adiabatic (A) and non-adiabatic (NA) effects.  The black solid line was carried out with typical electronic broadening, $T_1 = 1580$ K, which is large enough to wipe out the Kohn anomaly in linear response self-consistent force constants. The right-hand side graph shows a comparison between  $\alpha^{2}\bf F$s related to high adiabatic temperature ($T_1 = 1580$ K) with the  low non-adiabatic temperature that is fixed at the superconducting transition temperature, $T_0^{ {NA}} $= 26 K.  (d) Phonon dispersion as a function of electronic temperatures for the case +0.01   in both A and NA regimes.    $\alpha^{2}\bf F$s related to high adiabatic temperature ($T_1 = 1580$ K) and  low non-adiabatic  temperature, fixed at superconducting temperature ($T_0^{ {NA}} $= 65 K) are depicted in the graph of the right hand side. }
\label{Fsurface}
\end{figure}

The value of $T_c$s with the corresponding $N(\varepsilon_{F}$), $\lambda_{{tot}}$ and $\omega_{{log}}$ is shown as a function of the doping values in Table \ref{table1}. When non-adiabatic phonon dispersion is taken into account, there is almost  no difference between non-adiabatic low-temperature and high-temperature adiabatic phonons (see figures.~\ref{Fsurface}(c) and (d))  and their respective $T_c$s,  while $\lambda_{{tot}}$ and $\omega_{{log}}$ show a strong modulation as a function of temperature at low temperature considering adiabatic phonon dispersions (not shown here). It should be noted that $\alpha^2\bf F $ is calculated based on the high-temperature adiabatic approach, which is often a reasonable approximation. However, since there is a temperature-dependent modulation of the phonon dispersion due to the presence of the Kohn anomaly, the $\alpha^2\bf F$ must be evaluated at a specific temperature of our interest. In particular, the estimation of $T_c$ should be performed in the self-consistent way~\cite{Note3}. 

In P-doped cases, the $\lambda$ and $T_c$ are enhanced towards more substantial doping levels. There are two points corresponding to the P-dopings. First, a comparison between +0.01 and +0.03 doping levels shows that the $N(\varepsilon_{F})$ is enhanced $\sim\,$57$\%$, while the $\lambda$ is enhanced $\sim\,$36$\%$, which is a signature of an enhanced $\langle{{\bf g}}^2\rangle$ for doping level +0.01 in comparison with +0.03. Second, although $T_c$ is increased for deeper doping levels, such an increment of $\lambda$ leads to stronger phonon softening so that it yields a CDW phase for temperatures above related $T_c$ as long as one considers adiabatic phonons. In the following, we will carefully discuss the details of the CDW formation for doping values.

For the N-doped levels, both $\lambda$ and $N(\varepsilon_{F})$  are amplified towards the Lifshitz transition point which can occur at doping of $-$0.1. A comparison between $-0.05$ and $-0.1$ shows an enhancement of
84$\%$ and 170$\%$ for $N(\varepsilon_{F})$ and $\lambda$
respectively, indicating a stronger $\langle{{\bf g}}^2\rangle$ for doping level $-$0.1. Such a significant improvement of the $\langle{{\bf g}}^2\rangle$ is concurrent with the Lifshitz transition embracing larger phase space for electron-phonon scattering as illustrated in figure~\ref{Fsurface}(a). 

We develop our analysis by looking at the FS for various levels of doping. In figure~\ref{Fsurface}(a), the FS corresponding to doping $-$0.1 is depicted. As seen here, by increasing the doping level, the FS evolves its topology from six isolated drop-like pockets to dumbbell-shaped pockets that are merged into adjacent 1BZ: called the Lifshitz transition. Moreover, the pockets relevant to electron doping are located around $\sim0.5 \,\Gamma M$ and $\sim0.5\,\Gamma {M}'$ points in the 1BZ (see various high symmetry point labels in figure~\ref{Fsurface}(b)). This significantly restricts scattering processes into two types of large $\bf q$ vectors. The first is $|\bf{q}|\sim M$ including scatterings such as $ \, {M}_i \leftrightarrow {M}'_{j \neq i}$, $  {0.5}\,\Gamma {M}_i \leftrightarrow  {0.5}\,\Gamma {M}_{j \neq i}$. The second is $|\bf q| \sim\,K$ encompassing scattering like $  {0.5}\,\Gamma {M}_i \leftrightarrow  {0.5}\,\Gamma {M}'_i$ and $  {0.5}\,\Gamma {M}_i \leftrightarrow  {0.5}\,\Gamma {M}_{j\neq i}$.

Figure~\ref{Fsurface}(b) depicts the FS for a +0.01 doping including the pockets  located around $\sim{0.5}\Gamma K$ and $\sim{0.5}\Gamma{K}'$ points (below Lifshitz transition). In this case, the shape of the FS can remarkably contain the scattering processes with $ |\bf q|\sim\,K$ including scatterings like $ {0.5}\,\Gamma {K}_i \leftrightarrow  {0.5}\,\Gamma {K}'_i$, and $|\bf q|\sim\,M$ including scatterings, for instance, $  {0.5}\,\Gamma  {K}_i \leftrightarrow  {0.5}\,\Gamma {K}_{j\neq i}$. 
Thus in both cases, the system is expected to show a pronounced Kohn anomaly on these $\bf q$ vectors as long as tangible amounts of $\chi_0$ and/or $ \langle{{\bf g}}^2\rangle$ occur as will be expressed in the  following (see the origin of CDW instability).

Figure~\ref{Fsurface}(c) shows adiabatic and non-adiabatic phonon dispersions at high-temperature (1580 K or 0.01 Ry) and low temperature $T = 26$ K for N-doped level $-$0.1. While the phonon dispersion shows  softening as a function of temperatures, it does not show any CDW formation even for  temperatures as low as 2 K. Such a  character was repeated for other doping levels as mentioned in Table~\ref{table1}. Moreover, the Kohn anomaly displayed in the high symmetry point like the K  point has already been illustrated in the monolayer antimonene ~\cite{lugovskoi2018electron,PhysRevB.101.205412} and is predicted for other 2D buckled structures of the nitrogen family.
 A significant variation of the fourth branch (labeled by $E_g$ at the long wavelength limit) as a function of temperature results in a temperature-dependent $\alpha^2\bf F$ and $\lambda$: decreasing the temperature leads to enhancement of the $\lambda$. The adiabatic technique overestimates the enhancement; therefore, non-adiabatic phonons must be used for the evaluation of the $\alpha^2\bf F$ and the estimation of $T_c$.
Accordingly, if there is a strong temperature-dependent Kohn anomaly, it is preferable that one evaluates $T_c$ in the self-consistent approach, as shown in~\cite{Note3}. 
 
 Therefore, the most physical $\alpha^2\bf F $ is the one that is calculated based on non-adiabatic phonon dispersions at the corresponding temperature. In this way, as demonstrated in the figure~\ref{Fsurface}(c),  $\alpha^2\bf F $ is calculated based on non-adiabatic phonon modes at $T_0$ = 26 K leading to $\lambda$ = 2.11 and corresponding $T_c$ = 26 K. Note that if one uses high-temperature $\alpha^2\bf F$, the $\lambda$ becomes equal to 2.05 and the resulting $T_c$ = 22 K as reported in Table~\ref{table1}.  Furthermore, figure~\ref{Fsurface}(c) illustrates a Raman frequency shift at the zone center. The three optical branches of BLP consist of one $A_{1g}$ mode attributed to out-of-plane vibration and double degenerate $E_{g}$ modes referring  to in-plane displacements which both of them are Raman active and show  phonon frequency shift about 13 cm$^{-1}$  at room temperature compared with high-temperature adiabatic phonon calculations.

 
Figure~\ref{Fsurface}(d) shows both the adiabatic and non-adiabatic phonon dispersions as a function of temperature for the case +0.01. The phonon softening occurs at $\bf q\sim M$ and $\bf q\sim K$ towards temperature reduction accompanied by both adiabatic and non-adiabatic modes at $T = $ 105 and 65  K  referring to the $T_{  CDW}$ and $T_c$ (see Table~\ref{table1}). This is consistent with the FS topology as discussed before where the  softening occurs at $\bf q\sim M$ and $\bf q\sim K$. Note that the difference between adiabatic and non-adiabatic frequencies is significant for all aforesaid P-type doping  $+0.01, +0.02$   and $+0.03 $ due to encompassing large amounts for both the $N(\varepsilon_{\rm F}$) and  $ \langle{{\bf g}}^2\rangle$, while, as inferred from Table~\ref{table1}, this difference for two  N-type doping $-0.1$  and $-0.15$ is related to remarkable amount of $ \langle{{\bf g}}^2\rangle$;  essentially restating the importance of the  considering non-adiabatic effects.  The phonon softening is completely eliminated  by considering non-adiabatic phonon dispersions, which in turn postpone the CDW formation by further lowering the temperature.  In figure~\ref{Fsurface}(d), $\alpha^2\bf F$ is calculated at both high adiabatic temperature (1580 K) and low non-adiabatic temperature (65 K). Obviously, such a large reduction of the temperature does not significantly affect $\alpha^2\bf F$ as long as the non-adiabatic phonon dispersions are considered, so that 
 $T_c =$ 65 K can be slightly  upgraded  to $T_c^{ {NA}} =$ 67 K by using self-consistent method~\cite{Note3}. 
 Note that,  results in Table~\ref{table1} show that BLP exhibits a dynamically stable phase for the aforementioned N- and P-type doping levels at room temperature in the presence of the non-adiabatic phonon dispersion.



   \begin{figure}[!]
\centering
\includegraphics[scale=0.450,trim={0cm 0cm 0cm 0cm} ]{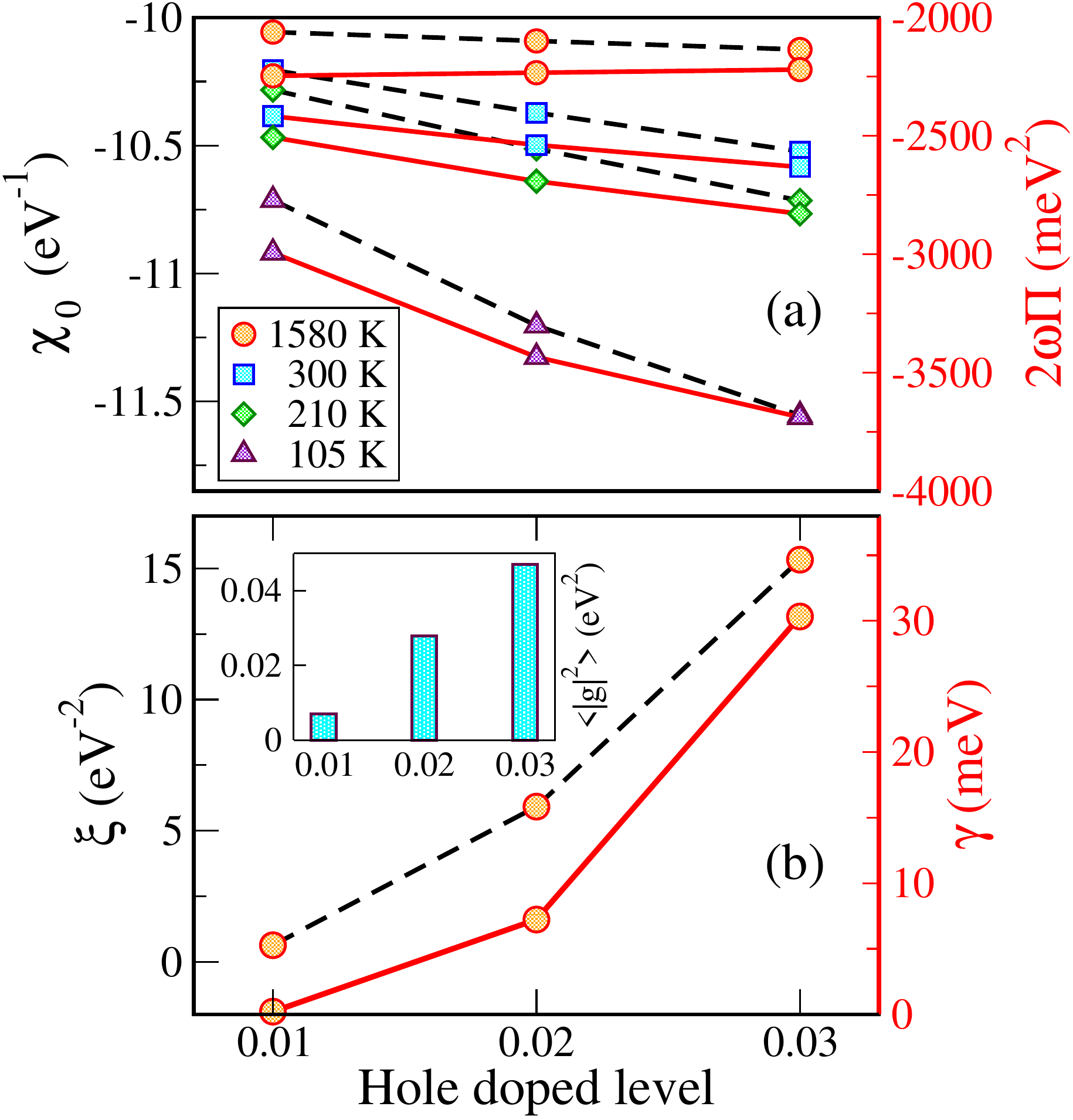}
\caption{  The principal factors to specify the origin of the  CDW instability as a function of temperature  at  $\textbf{q}_{{CDW}}= {M}$ for all P-doped levels of the single-layer  BLP in the adiabatic regime. (a) Electronic bare susceptibility, $\chi_0$, and  real part of the phonon self-energy, $2\omega \Pi$,(b)  nesting function, $\xi$,  phonon linewidth, $\gamma$. The splines connecting the points are guides to the eyes. The inset graph shows the average squared amounts of the electron-phonon strength for the three aforementioned P-doped levels.  }
\label{CDW-effective-hole}
\end{figure}

 As indicated in Table.~\ref{table1}, the entry into the CDW region occurs at temperature 105, 210 and 300 K for P-doping +0.01, +0.02 and +0.03 respectively as long as the adiabatic phonons is assumed. Hence, to perceive the underlying parameters which control the softening, the relevant quantities corresponding to $\textbf q_{{CDW}} = \,M$ are presented in figure~\ref{CDW-effective-hole} for diverse hole-doped levels and the above-mentioned temperatures. In figure~\ref{CDW-effective-hole}(a) the bare susceptibility (dashed line) together $2{\omega}{\Pi}$ (solid line) are depicted. The most significant modulation of the phonon energy belongs to the hole doping +0.03, which acquires the largest variation of $\chi_0$. The large variation of $\chi_0$ represents a result of larger nesting, $\xi$, at +0.03, shown in figure~\ref{CDW-effective-hole}(b), which successively is a result of enhanced $N(\varepsilon_F)$ together with the extension of the sharp edge of the pockets upon increasing hole dopings. Another reason for an enhanced softening for larger doping levels at $\textbf q_{{CDW}} = \,M$ is the fact that the averaged EPI strength, $\langle \textbf{g}^2\rangle_{\textbf {q}\nu}= \frac{\gamma_{\textbf {q}\nu}}{2\pi\omega_{\textbf {q}\nu}\xi_{\bf q}}$, extracted from relation between nesting and the phonon  linewidth ($\gamma$) mentioned in the supplement, is reasonably larger for deeper doping levels (+0.02 and +0.03) as illustrated in the inset of figure~\ref{CDW-effective-hole}(b). The combination of the large nesting together with an enhanced EPI for deeper doping levels results in the persistence of the Kohn anomaly as long as the adiabatic phonon dispersion is considered. Therefore, considering the  adiabatic approach, incorrectly a CDW phase will  prevail and make an obstacle to achieve  superconducting critical temperatures.


    \begin{figure}[t]
\centering
\includegraphics[scale=0.41,trim={-6cm 0cm 0cm 0cm} ]{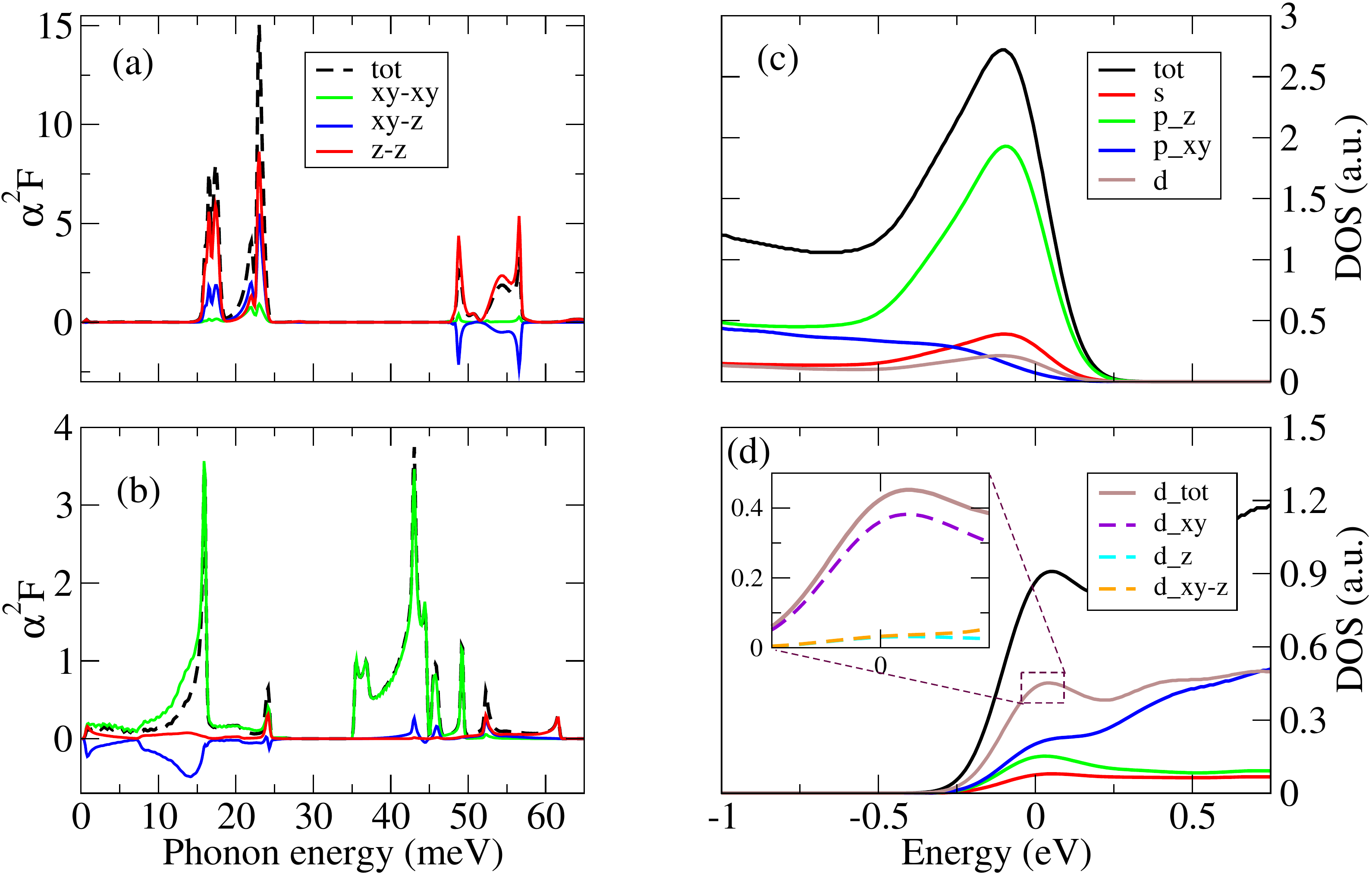}
\caption{Projected $\alpha^{2}\bf F$ and electronic DOS along the in-plane and out-of-plane directions. (a) and (b) display projected $\alpha^{2}\bf F$ for doping levels for two cases +0.01 and -0.1 while (c) and (d) illustrate electronic DOS for two doping levels +0.01 and -0.1,  respectively. The Fermi levels are set to zero in DOS's graphs. The inset of the (d) shows distinct contributions related to the d orbital projected in two principal directions.}
\label{proj.a2f.pdos}
\end{figure}

To find out what kind of phonons and electrons are coupled,  we evaluate the projected $\alpha^2 \bf F$ into in-plane ($xy$) and out-of-plane ($z$) directions (see  details in the supplement) 
for +0.01 and -0.1 doping levels as shown in figures.~\ref{proj.a2f.pdos}(a) and \ref{proj.a2f.pdos}(b), respectively.
For +0.01 case, the resulting projected $\lambda$s are $\lambda_{xy,xy} =0.3,\,\, \lambda_{xy,z} = 0.86$ and $\lambda_{z,z}=2.62$ indicating the most contribution is related to  out-of-plane displacements of the phonons. 
This is equally consistent with the fact that the $p_z$ orbital allows for a larger contribution in the wave functions of the valence band for these particular doping levels. This is more explicitly demonstrated  in figure~\ref{proj.a2f.pdos}(c), where the projected electronic DOS corresponding to a doping level +0.01 is shown; the out-of-plane deformation of phonons is coupled to electrons with a tangible contribution from the p$_z$ orbitals.

The same analysis corresponding to doping level $-$0.1 shows that $\lambda_{xy,xy}=2.55, \,\, \lambda_{xy,z}=-1.04$ and $ \lambda_{z,z}= 0.54$. The largest contribution arises from the in-plane displacements of phonons. This is also consistent with the fact that in-plane components of d and p orbitals have the largest contribution to the wave functions of the conduction band for these types of doping levels. This is more explicitly shown in figure~\ref{proj.a2f.pdos}(d), where the projected DOS corresponding to doping level $-$0.1 is considered. In addition, distinct components of the d orbital dissociated in two principal directions are illustrated in the inset of the figure~\ref{proj.a2f.pdos}(d).

Although the anisotropy existing in the systems can be eliminated by their intrinsic impurity, in several 2D materials such as graphene~\cite{PhysRevB.90.014518} and MgB$_2$~\cite{PhysRevB.66.020513}, a comparison between $T_c$ extracted from the Allen-Dynes formula  and experiment reveals the importance of  anisotropic effects and, moreover, an underestimate  of $T_c$ is obtained in the absence of an anisotropic formalism.

\begin{figure}[t]
\centering
\includegraphics[scale=0.43,trim={0cm 0cm 0cm 0cm} ]{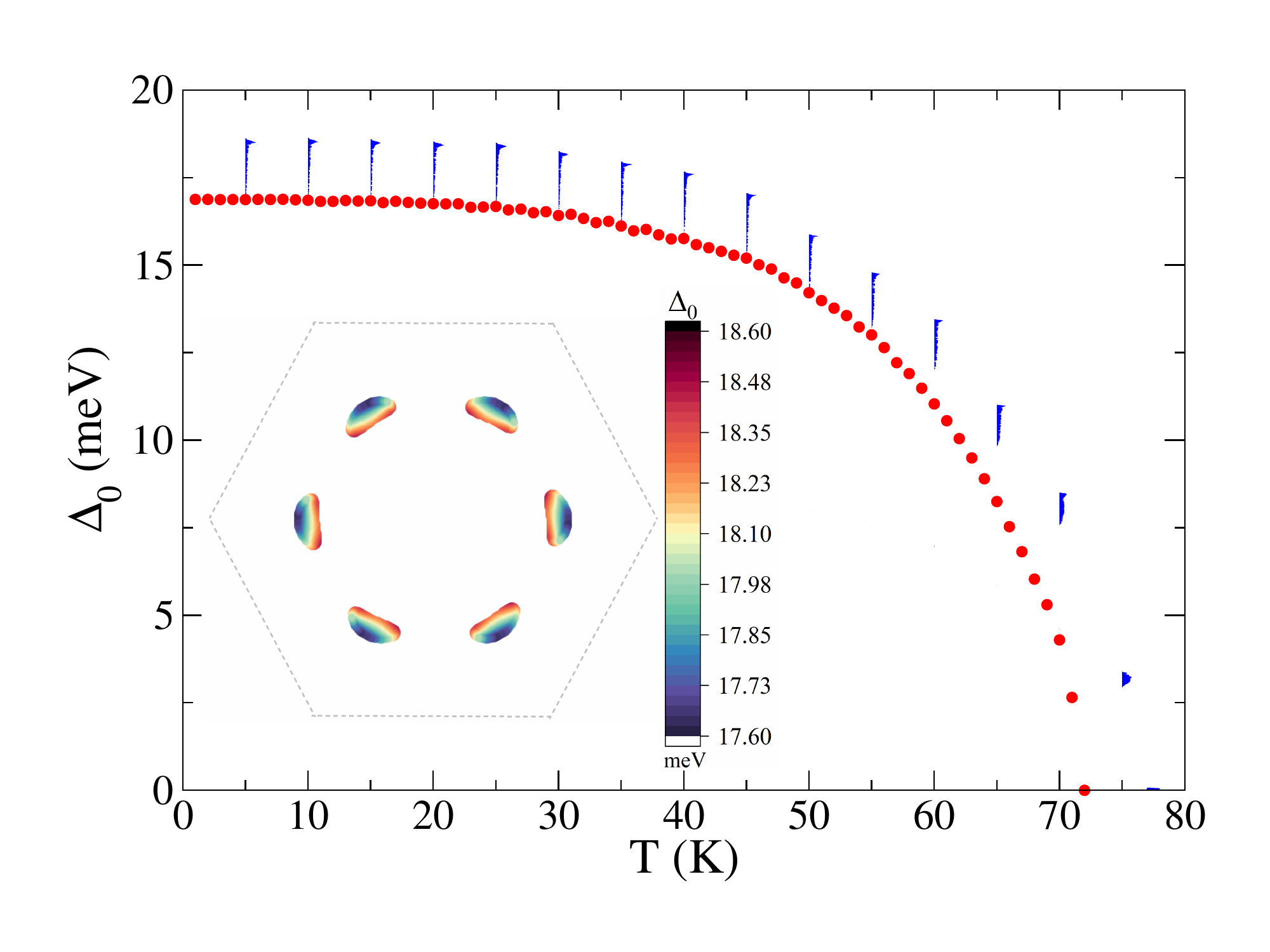}
\caption{The variation of the isotropic superconducting gap, $\Delta_0$ (red circles), as well as energy distribution of anisotropic superconducting gap, $\Delta_{0\bf K}$ (blue) as a function of temperature for the case $0.01$ of monolayer BLP. The effective Coulomb parameter is set to $\mu_c^* $ = 0.1. The inset shows  the calculated momentum-resolved superconducting gap at 5 K on the FS.}
\label{Gap-hole-0.01}
\end{figure}

As an example of P-doping, figure~\ref{Gap-hole-0.01} shows the superconducting gap ($\Delta_0$) as a function of temperature at the Fermi energy for the case +0.01 using isotropic (red circles)  {\it ab initio} Migdal-Eliashberg (ME) theory when $\mu_c^* $ = 0.1 (for  more details on choosing $\mu_c^* $, see the supplement). The $\Delta_0$ at low temperature is found to be $\Delta_0 \sim$  17.5 meV. In this framework, the  $T_c$ is defined  as the temperature at which the  $\Delta_0$ tends to quench which explicitly leads to  $T_c$ = 72 K. As we expected, a remarkable deviation of the BCS universality ($2\Delta_0/T_c = 3.53$) is observable owing to a strong EPC. 
 
\begin{figure}[t]
\centering
\includegraphics[scale=0.350,trim={0cm 0cm 0cm 0cm} ]{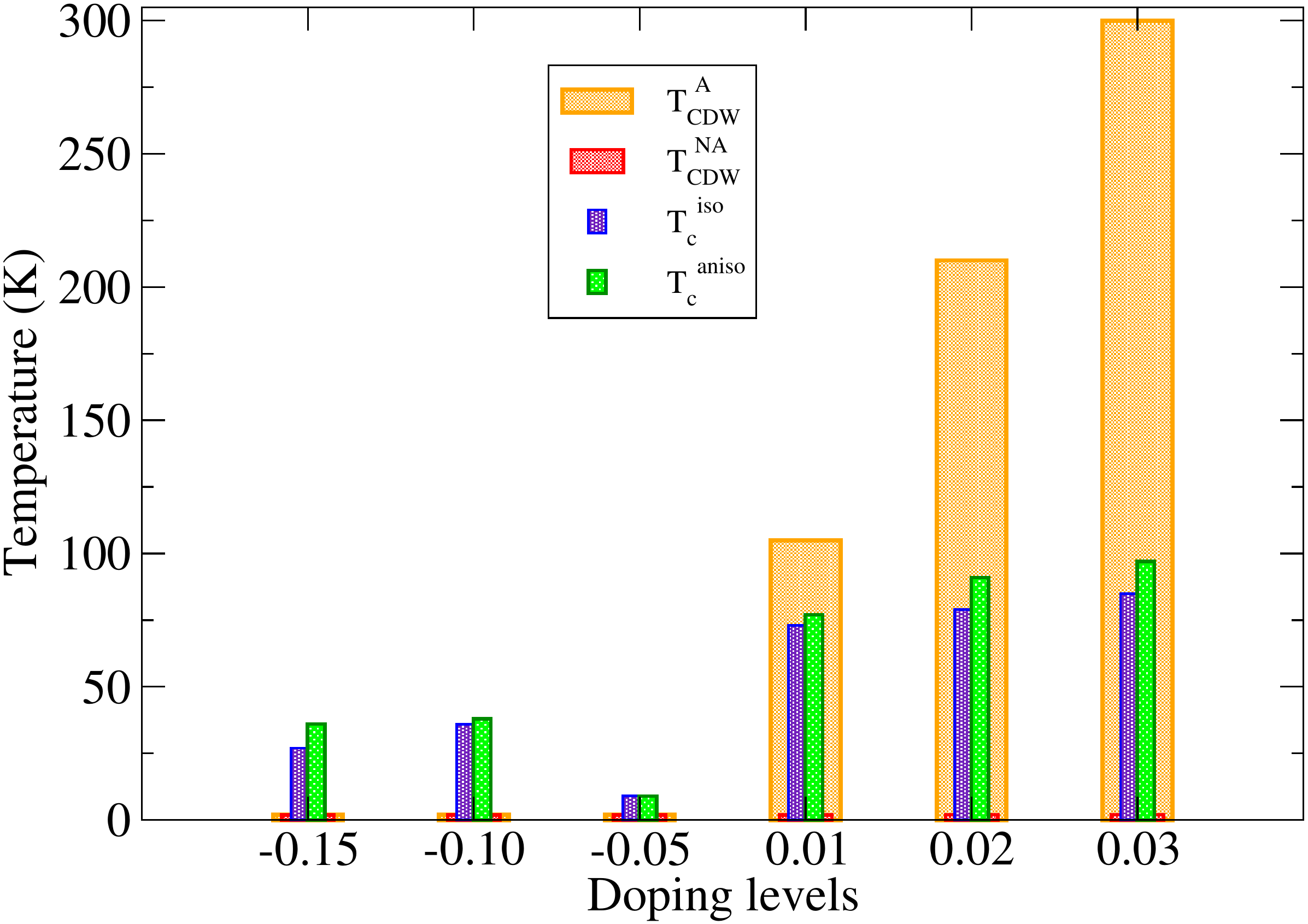}
\caption{The competition between superconducting transition temperature extracted by the ME theory and transition temperature to CDW phase within both adiabatic and non-adiabatic regimes for various doping levels. }
\label{ME-iso-anaiso-t-tcdw}
\end{figure}

 Furthermore, figure~\ref{Gap-hole-0.01} shows the energy distribution of $\textbf k$-dependence of the $\Delta_0$ on the FS for various temperatures, within anisotropic ME theory (blue). For $T< 60$ K, the distribution is practically temperature independent and shows a narrow maximum around 18.5 meV accompany with a vast flat area approximately in the range of 16.6-18 meV.  The inset figure explains that the high energy narrow maximum region is related to an area around the $0.5\Gamma K$ point which is related to the FS (see figure~\ref{Fsurface}(b)). As the anisotropic ME theory is used, this region significantly improves $T_c$, whereas the flat part corresponding to the lower energy levels has a less impact. 
Comparing the isotropic with anisotropic ME theory, shows an enhancement about $\Delta T_c = 5$ K, leading to a $T_c=77$ K for the anisotropic case. The $T_c$ for the rest of the P-dopings can be found in figure~\ref{ME-iso-anaiso-t-tcdw} for both isotropic and anisotropic cases.  In addition, more details indicated that the larger FS sheet around the $0.5\Gamma K$, the wider spread in the energies of the $\Delta_0$. Therefore, the more remarkable difference between isotropic and anisotropic $T_c$ can be found in Tables S2 and S3).

\begin{figure}[t]
\centering
\includegraphics[scale=0.43,trim={0cm 0cm 0cm 0cm} ]{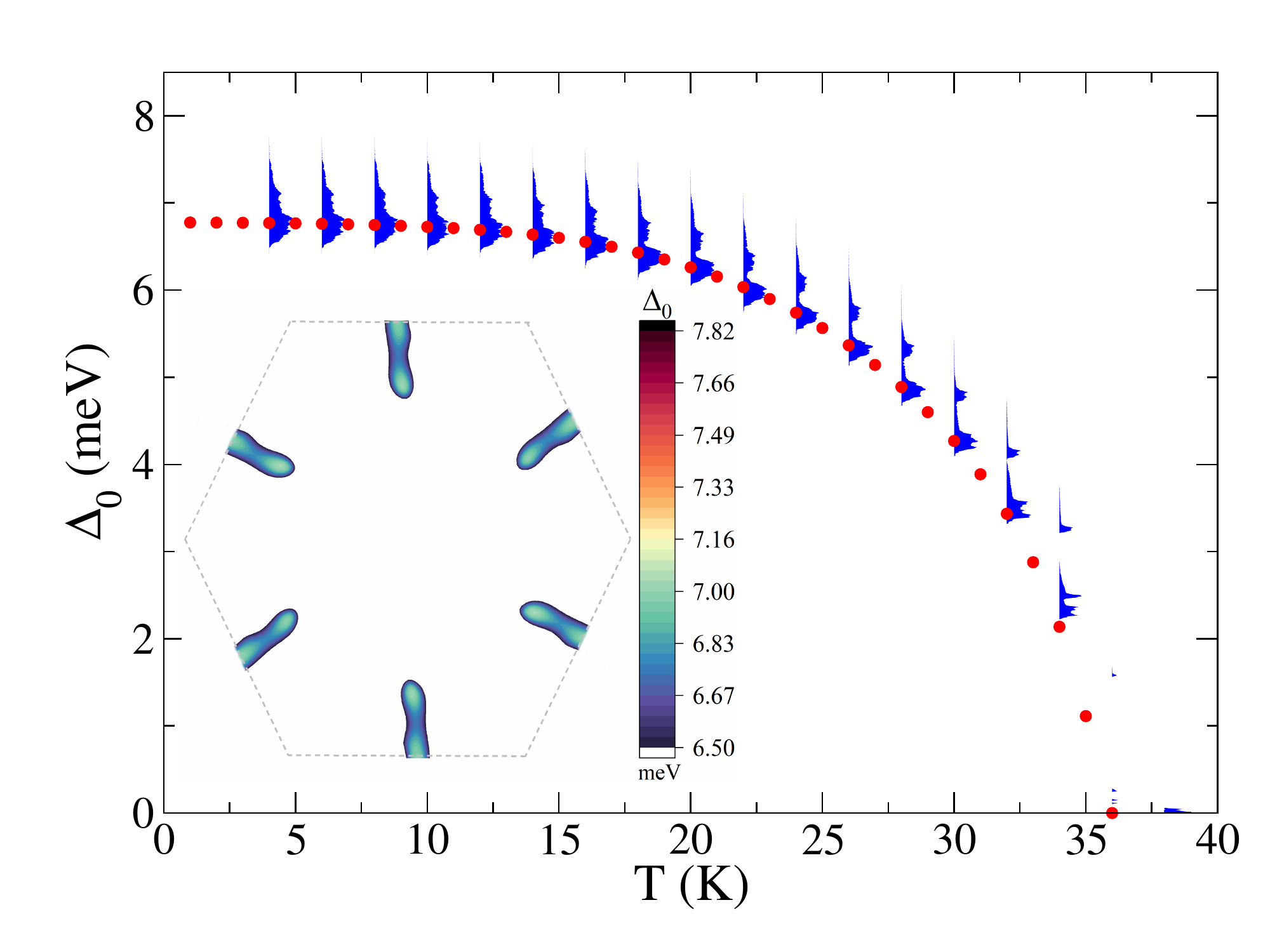}
\caption{The variation of the isotropic superconducting gap, $\Delta_0$ (red circles), as well as energy distribution of anisotropic superconducting gap, $\Delta_{0\bf K}$ (blue) as a function of temperature for case $-0.1$ of monolayer BLP. The effective Coulomb parameter is set to $\mu_c^* $ = 0.1. The inset shows  the calculated momentum-resolved superconducting gap at 4 K on the FS.}
\label{anis0-el-0.1}
\end{figure}

In N-doped cases, we scrutinize the variation of the anisotropic $\Delta_0$ with respect to different temperatures across the FS for the case $-$0.1 plotted in figure~\ref{anis0-el-0.1}. This result shows the  $\Delta_0 \sim$7.5 meV accompany by $T_c$ = 38 K  implying  a notable deviation from the BCS theory. The role of the FS to determine the alteration of the $\Delta_0$ as a function of temperatures specifies the existence of two distinct parts as depicted in the inset of figure~\ref{anis0-el-0.1}. The first, at the lower energy ($\Delta_0 <$ 7 meV), includes a wide maximum part encompassing the Lifshitz point, while the second is related to a drop-like area located in the higher energy levels ($\Delta_0 \gtrsim$ 7 meV) as illustrated in the inset of figure~\ref{anis0-el-0.1}. At this doping level, the isotropic and anisotropic $T_c$’s are approximately analogous $\Delta T_c = 2$ K. 
It should be pointed out that in the deeper doping level with more anisotropy, more difference of $T_c$ is expectable  (Tables S.2 and S.3). 

Figure~\ref{ME-iso-anaiso-t-tcdw} illustrates results related to competition between the CDW and superconducting phases (in both adiabatic and non-adiabatic regimes) by considering both isotropic and anisotropic ME theory. The results show that the presence of non-adiabatic effects makes an obstacle to entrance into the unstable CDW phase for the aforesaid P-doped levels. Moreover, in the case of P-doping, $T_c$ increases for higher doping levels so that maximum  $T_c$'s are calculated about 97 and 91 K for doping +0.03 and +0.02 in the anisotropic ME formalism, respectively. While in the case of electron doping,  crossing from case $-$0.05 ($T_c =$ 9 K), firstly, an enhancement at the $T_c =$ 38 K and next decreasing of that is visible.   
Owing to the fact that various factors like underlying substrate and dimensionality alter dielectric screening and therefore strongly affect $\mu^*_c$, the results related to  $T_c$ arising from other phenomenological amounts of $\mu^*_c$ from 0.1 to 0.2  were tabulated  in the section S2 of the supplement as well. Those results indicate that, for instance, increasing in $\mu^*_c$ till 0.2 yields  decreasing in $T_c$ about 23$\%$ for case +0.03 and  37$\%$ for the case $-$0.10.

\section{Conclusions}\label{sec:conclusion}

In summary, we have studied the superconducting properties of single-layer blue phosphorene in the hole (P)- and electron (N)-doped cases. By assuming non-adiabatic phonon dispersion for the P-doping regime, we have reached a maximum critical temperature $T_c$ = 97 K for the +0.03 case when the anisotropic Migdal-Eliashberg theory was used. Considering only the adiabatic phonon dispersions for all P-doped levels leads to a strong softening at temperatures well above the corresponding estimated $T_c$, while the presence of the non-adiabatic phonon dispersions reduces  $T_{{CDW}}$. For the N-doped levels, based on non-adiabatic phonon dispersion, a maximum of the electron-phonon coupling constant, $\lambda=2.11$, was obtained for an electron doping of $-0.1$, resulting in a maximum of $T_c$ = 38 K via  the anisotropic Migdal-Eliashberg theory. Projected $\alpha^{2}\bf F$ and electronic density of state reveal that remarkably $\lambda$'s are attributed to the coupling between in-plane/out-of-plane phonons with in-plane/out-of-plane electronic states in the N/P type doped levels, respectively. Moreover, the results well express that a monolayer BLP is dynamically stable for both aforesaid N- and P-type doping at room temperature in the presence of the non-adiabatic phonon dispersion. Finally, we made clear that the absence of non-adiabatic effects prevents a thorough examination of the rivalry between CDW and the superconducting phases.

\ack
R. A thanks for the support from the Australian Research Council Centre of Excellence in Future Low-Energy Electronics Technologies (project number CE170100039).


\section*{References}
\bibliographystyle{iopart-num}
\bibliography{draft1.bib}



\end{document}